\documentclass[preprint,12pt]{elsarticle}

 \usepackage{graphics}

\usepackage{amssymb}

\biboptions{super}

\journal{Physics Letter A}

\begin{document}

\begin{frontmatter}



\title{p-wave superconductors in dilaton gravity}


\author{ZhongYing Fan}

\address{Department of Physics, Beijing Normal University, Beijing 100875, China}

\begin{abstract}
  In this paper, we study peculiar properties of p-wave superconductors in dilaton gravity. The scale invariance of the bulk geometry is effectively broken due to the existence of dilaton. By coupling the dilaton to the non-Abelian gauge field, i.e., $-\frac14 e^{-\beta \Phi} F^a_{\mu\nu}F^{a\mu\nu}$, we find that the dissipative conductivity of the normal phase decreases and approaches zero at the zero frequency as $\beta$ increases. Intuitively, the system behaves more and more like an insulator. When the hairy solution is turned on, the system crosses a critical point to the superconducting phase. We find that the critical chemical potential decreases with the increasing of $\beta$ and the maximum height of the conductivity is suppressed gradually which are consistent with our intuition for insulator/supercondutor transition.
\end{abstract}


\end{frontmatter}

\section{Introduction}
In recent years, AdS/CFT correspondence has been widely used to investigate strongly correlated condensed-matter physics. One of remarkable achievements is holographic model of superconductors\cite{1,2,3,4}. The superconductors described in the weakly coupled gravitational background are dual to those strongly coupled superconductors in the boundary theory. It is an excellent example to show the power of gauge/gravity duality since the strongly coupled superconductors still lacks an effective description in condensed-matter physics. In contrast, the AdS/CFT correspondence provides an elegant and systematic procedure to study the peculiar properties of superconductors. It is also possible to develop a deep understanding on various mysteries such as the paring mechanism and enhanced critical temperature of superconductors.

Since the boundary theory is in general not scale invariant below some dynamical scale, it is more important to study non-relativistic holography with scale invariance broken in the bulk. In this paper, we introduce a dilaton field to effectively break the scale invariance of the bulk geometry\cite{5,6,7,8}. We then move on to discuss p-wave superconductors using SU(2) non-Abelian  gauge field\cite{4}. By rescaling the gauge coupling constant as $1/g_F\rightarrow e^{-\beta \Phi}/g_F$, we couple the dilaton to the gauge field. The parameter $\beta$ is chosen to be positive. As $\beta$ increases, we find that the real part of the conductivity monotonically decreases. The DC conductivity defined at the zero frequency approaches zero and vanishes when $\beta$ exceeds some critical value. This indicates the characteristics of insulators. With the increasing of $\beta$, the system behaves more and more like an insulator in the normal phase. When the chemical potential crosses a critical point, the system begins to superconduct. The order parameter defined by $A_x^3$ condenses, leading to a hairy near the horizon of the black hole and a gap is opened in the real part of the conductivity. Moreover, there exists a Drude-like structure at some frequency close to the zero frequency point in the conductivities along the x-direction. This is similar to the result published in ref.\cite{4}. When $\beta$ increases, we find that the critical chemical potential in general decreases and the maximum height of the dissipative conductivity is suppressed, consistent with our intuition for insulators.

The remainder of this paper is organized as follows: In section 2, we briefly introduce the gravity solution in Einstein-dilaton model. In section 3, we present the holographic model of p-wave superconductors coupled to the dilaton and derive the equations of motion (Eoms). We numerically find that the order parameter condenses when the chemical potential crosses a critical value. In section 4, we investigate the fluctuation modes of Yang-Mills theory. In the end, we present a short conclusion in section 5.

\section{Gravity solution}
Let's consider the standard Einstein-dilaton action:
\begin{equation} S_{Grav}=\frac{1}{2\kappa^2}\int \mathrm{d}^4x\sqrt{-g}\ [R-(\partial\Phi)^2-V(\Phi)],   \label{1}\end{equation}
where $\kappa^2$ is the Newton constant, $R$ is the Ricci scalar, $\Phi$ is a real scalar field (the dilaton) and the AdS radius has been set to 1. In order to obtain the dilaton black hole solutions, the potential $V(\Phi)$ has to be chosen appropriately. It reads\cite{5}:
\begin{equation} V(\Phi)=\sum_{i=1}^6 V_i e^{-\delta_i\Phi}, \end{equation}
where $V_i$ and $\delta_i$ are constants. The general static black hole solutions with asymptotical $\mathrm{AdS}_4$ geometry have the following form:
\begin{equation} ds^2=W(r)(-f(r)dt^2+\frac{dr^2}{f(r)}+dx^2+dy^2),\quad \Phi=\Phi(r). \end{equation}
The functions $W(r)$, $f(r)$ and $\Phi(r)$ are given by
\begin{equation} W(r)=\frac{\nu^2(1+r)^{\nu-1}}{[(1+r)^\nu-1]^2},\quad \Phi(r)=\pm\sqrt{\frac{\nu^2-1}{2}}\log{(1+r)}, \end{equation}
\begin{equation} f(r)=1+\frac{3}{r_s^3} \{ \frac{\nu^2}{4-\nu^2}+(1+r)^2[1-\frac{(1+r)^\nu}{2+\nu}-\frac{(1+r)^{-\nu}}{2-\nu}]  \}.  \end{equation}
where $\nu$ is a constant, satisfying $\nu\geq 1$, $r_s$ is also a constant. The constants $V_i$ and $\delta_i$ can be expressed in terms of $\nu$ and $r_s$. Since the potential of the dilaton is symmetric under $\Phi\rightarrow -\Phi$\cite{5}. From now on, we focus on $\Phi\geq 0$.

The black hole horizon is defined by $f(r_h)=0$ and $r_h$ is determined by $r_s$ and $\nu$. A special case is $\nu= 1$. The dilaton vanishes and the dilaton black hole with scale invariance broken is reduced to the standard scale invariant Schwarschilld black hole with $W(r)=1/r^2$, $f(r)=1-(r/r_s)^3$. In this case $r_h=r_s$. Therefore, the scale invariance broken of the bulk geometry can be measured by the difference $\nu-1$\footnote{In fact, this difference $\nu-1$ is proportional to the hyperscaling violation $\theta$ in the zero temperature limit, which directly measures how strongly the scale invariance is broken.}. In the next sections, we will set $r_s=1$ for convenience.

\section{P-wave superconductors coupled to a dilaton}
In order to investigate p-wave superconductors in holography, we start from SU(2) action
\begin{equation} S_{Higg}=-\frac{1}{4g_F^2}\int \mathrm{d}^4x\sqrt{-g} e^{-\beta\Phi} F^a_{\mu\nu}F^{a\mu\nu}. \end{equation}
where $F^a_{\mu\nu}=\partial_\mu A^a_\nu-\partial_\nu A^a_\mu+\epsilon^{abc}A^a_\mu A^a_\nu$, $A^a_\mu$ is the gauge potential, $\epsilon^{abc}$ is the structure constant of the Lie group. The dilaton has been coupled to the gauge field and rescale the effective gauge coupling constant. In the following, we will not consider the backreaction effect of the gauge field on the Einstein-dilaton sector by taking the probe limit $\kappa^2\rightarrow 0$. The equation of motion is obtained by variation of the action with respect to the gauge potential
\begin{equation} \frac{1}{\sqrt{-g}e^{-\beta\Phi}}\partial_\nu(\sqrt{-g}e^{-\beta\Phi}F^{a\nu\mu})+\epsilon^{abc}A^b_{\nu}F^{c\nu\mu}=0.\label{eom1} \end{equation}
To discuss the p-wave superconductors, we consider the following ansatz
\begin{equation} A=\varphi(r)\tau^3 \mathrm{d}t+\psi(r)\tau^1\mathrm{d}x. \end{equation}
It is straightforward to derive equation of motions for $\varphi(r)$ and $\psi(r)$
\begin{equation} \varphi''-\beta\Phi'\varphi'-\frac{\psi^2}{f}\varphi=0,\label{eom2} \end{equation}
\begin{equation} \psi''+(\frac{f'}{f}-\beta\Phi')\psi'+\frac{\varphi^2}{f^2}\psi=0. \end{equation}
where prime denotes $d/dr$. Recall that the U(1) subgroup of SU(2) generated by $\tau^3$ is specified to be electromagnetic. The function $\varphi(r)$ is interpreted as the time component of the U(1) field while $\psi(r)$ plays the role of order parameters. It is more convenient to work by rescaling the radial coordinate as $r\rightarrow r/r_h$. The above equations of motion remain unchanged with the fields rescaled as $\varphi\rightarrow \varphi r_h$, $\psi\rightarrow \psi r_h$. Notice that the location of the horizon has been scaled to unity.

To solve above equations of motion, we need to impose proper boundary conditions at the horizon
\begin{equation}\varphi(r)=\varphi_1 (r-1)+... \label{eom3} \end{equation}
\begin{equation} \psi(r)=\psi_0+\psi_1 (r-1)+... \label{eom4}\end{equation}
In the asymptotic limit, the fields are expanded as
\begin{equation} \varphi(r)=\mu-\rho r+... \label{eom5}\end{equation}
\begin{equation} \psi(r)=\psi^{(0)}+\psi^{(1)} r+... \label{eom6}\end{equation}

\begin{figure}
 \centering
  \includegraphics[width=7.5cm]{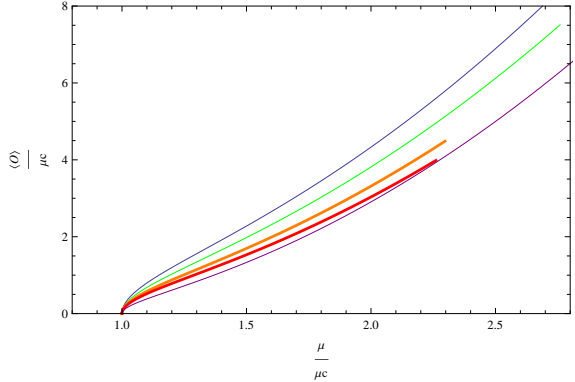}\\
  \caption{The vector condensate of order parameters for $\nu=1.4$. $\beta=0$ (blue), $\beta=1.5$ (green), $\beta=3$ (orange), $\beta=4$ (red), $\beta=6$ (purple).}\label{fig1}
\end{figure}

where dots ``..." denote higher order terms. The physical quantities such as chemical potential $\mu$, charge density $\rho$ and the condensate of the order parameter $\langle O \rangle$ of the dual field theory can be directly read off from these asymptotic behaviors. In order to obtain a stable theory with free source, the non-normalizable mode of $\psi(r)$ should be set to zero, i.e., $\psi^{(0)}=0$. This condition is realized by choosing $\varphi_1$ as the shooting parameter in numerics.

\begin{figure}
\centering
  \includegraphics[width=7.5cm]{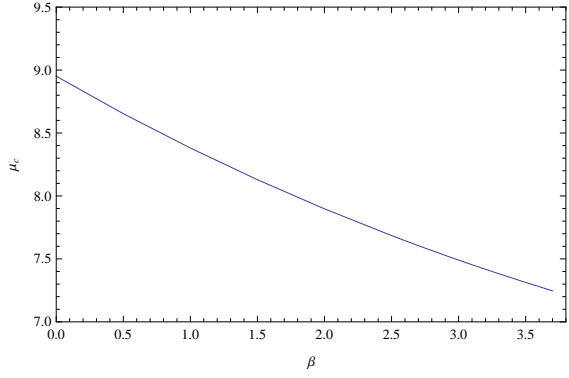}\\
  \caption{The critical chemical potential as a function of $\beta$ for $\nu=1.4$.}\label{fig2}
\end{figure}

From fig.\ref{fig1}, we find that the order parameter condenses when the chemical potential exceeds some critical value for various value of parameter $\beta$. As $\beta$ increases, the vector condensate measured in units of critical chemical potential decreases. Certainly, the critical chemical potential varies with $\beta$. We find that $\mu_c=8.9511,\ \mathrm{for}\ \beta=0$, $\mu_c=8.1291,\ \mathrm{for}\ \beta=1.5$, $\mu_c=7.4912,\ \mathrm{for}\ \beta=3$, $\mu_c=6.9825,\ \mathrm{for}\ \beta=4$ and $\mu_c=3.3821,\ \mathrm{for}\ \beta=6$, respectively. In fig.\ref{fig2}, we plot the critical chemical potential as a function of $\beta$. In general, $\mu_c$ decreases as $\beta$ increases. One can expect that when $\beta$ is sufficiently large, the critical chemical potential will approach to a small value, implying that the system has become stably like an insulator.

\section{Conductivity}
\subsection{Eoms and boundary conditions}
In order to analyse the electromagnetic perturbations of the U(1) subgroup of SU(2) generated by $\tau^3$, it is sufficient to obtain consistent linearized equations by considering fluctuation modes as follows
\begin{equation} a=e^{-i\omega t}[a_t^1(r) \tau^1 dt+a_t^2(r) \tau^2 dt+a_x^3(r) \tau^3 dx+a_y^3(r) \tau^3 dy]. \label{eom7} \end{equation}
From linearized Yang-Mills equations, we derive four second order differential equations
\begin{equation}{a_y^3}''+(\frac{f'}{f}-\beta \Phi'){a_y^3}'+(\frac{\omega^2}{f^2}-\frac{\psi^2}{f})a_y^3=0, \label{eom8} \end{equation}
\begin{equation}{a_x^3}''+(\frac{f'}{f}-\beta \Phi'){a_x^3}'+\frac{1}{f^2}(\omega^2 a_x^3-\varphi\psi a_t^1-i\omega \psi a_t^2)=0,  \label{eom9}\end{equation}
\begin{equation}{a_t^1}''-\beta \Phi'{a_t^1}'+\frac{\varphi\psi}{f}a_x^3=0,  \label{eom10}\end{equation}
\begin{equation}{a_t^2}''-\beta\Phi'{a_t^2}'-\frac{\psi}{f}(\psi a_t^2+i\omega a_x^3)=0, \label{eom11}\end{equation}
and two first order constraints
\begin{equation}i\omega {a_t^1}'+\varphi {a_t^2}'-\varphi'a_t^2=0,  \label{eom12}\end{equation}
\begin{equation}-i\omega {a_t^2}'+\varphi {a_t^1}'-\varphi' a_t^1+f(\psi {a_x^3}'-\psi' a_x^3)=0.  \label{eom13}\end{equation}
Notice that $a_y^3$ decouples from the other modes. Since it behaves identical to the fluctuation mode of s-wave superconductors except the slightly difference of the last term, the conductivity along the y-direction exhibits similar properties of s-wave case, which is less interesting in this paper. We will focus on discussing the coupled modes $\{a_x^3,\ a_t^1,\ a_t^2\}$ and the resulting x-direction conductivity $\sigma_{xx}$.
To extract retarded correlator, we need to impose infalling condition at the horizon for the coupled modes
\begin{equation} a_x^3=(r-1)^{-i\omega/\delta}[1+a_x^{3(1)}(r-1)+a_x^{3(2)}(r-1)^2+...], \label{eom14}\end{equation}
\begin{equation} a_t^1=(r-1)^{-i\omega/\delta}[a_t^{1(1)}(r-1)+a_t^{1(2)}(r-1)^2+...], \label{eom15}\end{equation}
\begin{equation} a_t^2=(r-1)^{-i\omega/\delta}[a_t^{2(1)}(r-1)+a_t^{2(2)}(r-1)^2+...], \label{eom16}\end{equation}
where, $\delta=|f'(1)|$. Integrating out to the boundary, the modes are expanded as
\begin{equation}a_x^3=A_x^{3(0)}+A_x^{3(1)}r+...  \label{eom17} \end{equation}
\begin{equation}a_t^1=A_t^{1(0)}+A_t^{1(1)}r+...  \label{eom18} \end{equation}
\begin{equation}a_t^2=A_t^{2(0)}+A_t^{2(1)}r+...  \label{eom19} \end{equation}
Since the conductivity $\sigma_{xx}$ is a physical quantity at the boundary, we need to do an infinitesimal SU(2) gauge transformation to construct a gauge invariant potential $\tilde{a}_x^3$ from the coupled modes $\{a_x^3,\ a_t^1,\ a_t^2\}$. The details were presented in ref.\cite{4}. Here, we write down the final result directly
\begin{equation} \tilde{a}_x^3=a_x^3+\frac{i\omega a_t^2+\varphi a_t^1}{\varphi^2-\omega^2}\psi. \label{eom20}\end{equation}
Near the boundary, it behaves as
\begin{equation} \tilde{a}_x^3=A_x^{3(0)}+[A_x^{3(1)}+\frac{i\omega A_t^{2(0)}+\mu A_t^{1(0)}}{\mu^2-\omega^2}\ \psi^{(1)}]\ r+... \label{eom21} \end{equation}
The conductivity $\sigma_{xx}$ can be directly read off from above expansion coefficients as follows
\begin{equation} \sigma_{xx}=\frac{1}{i\omega A_x^{3(0)}}[A_x^{3(1)}+\frac{i\omega A_t^{2(0)}+\mu A_t^{1(0)}}{\mu^2-\omega^2}\ \psi^{(1)}].\label{eom22} \end{equation}

\subsection{Conductivity in the normal phase}
In the normal phase, $\psi\equiv0$, the fluctuation mode $a_x^3$ decouples from $a_t^1$ and $a_t^2$ and behave identical to $a_y^3$, resulting to $\sigma_{xx}=\sigma_{yy}$. We will drop the subscript of the conductivity in this subsection. The decoupled mode satisfies
\begin{equation}{a_x^3}''+(\frac{f'}{f}-\beta \Phi'){a_x^3}'+\frac{\omega^2}{f^2} a_x^3=0  \label{eom23}\end{equation}
By imposing infalling boundary condition at the horizon, the conductivity can be read off from the far field expansion coefficients as well as the broken phase. \begin{figure}
\centering
  \includegraphics[width=6.5cm]{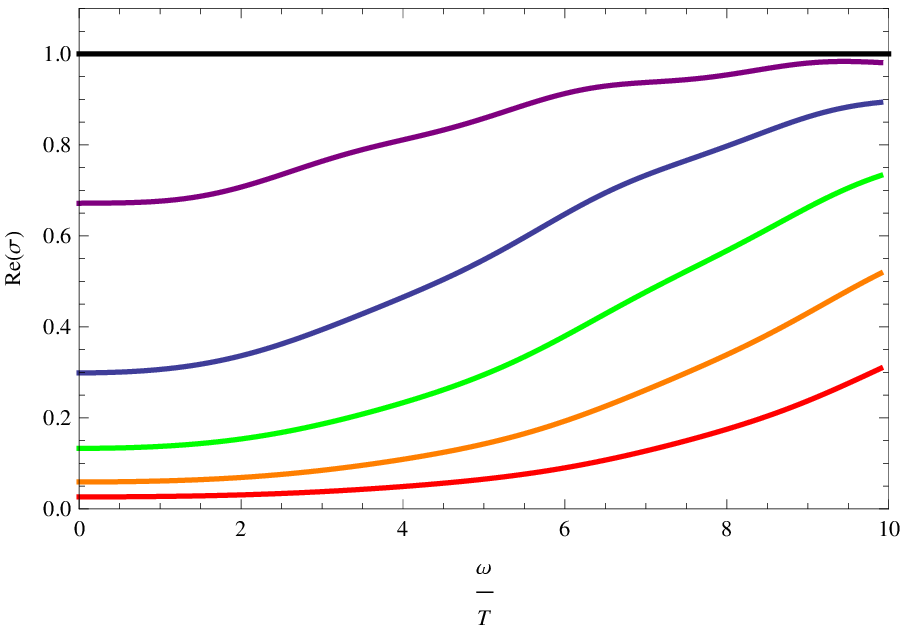}
  \includegraphics[width=6.5cm]{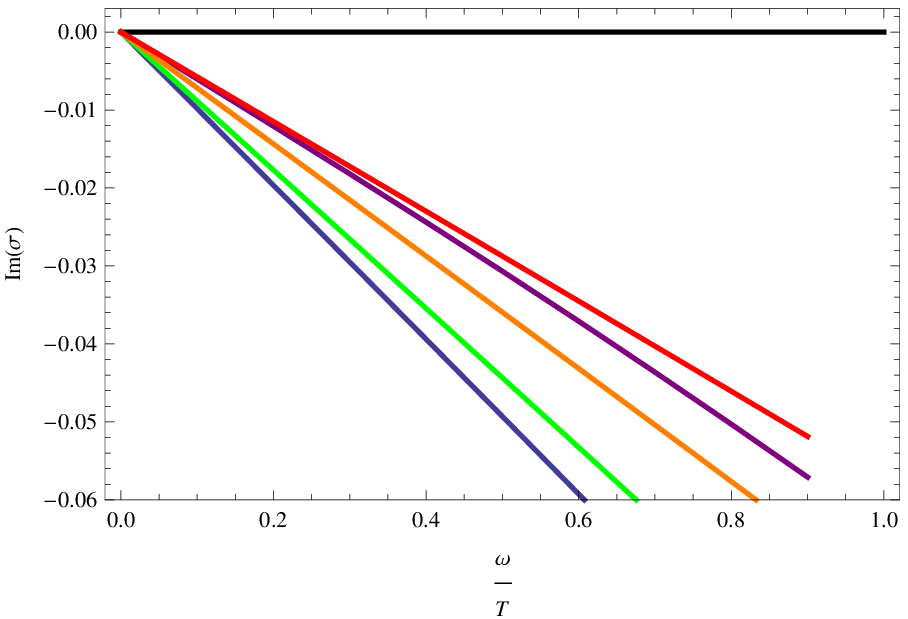}
  \caption{The conductivity of the normal phase for various parameter $\beta$. The left plot is the real part while the right plot is the imaginary part. $\beta=0$ (black), $\beta=1$ (purple), $\beta=3$ (blue), $\beta=5$ (green), $\beta=7$ (orange), $\beta=9$ (red).}\label{fig3}
\end{figure}
In fig.\ref{fig3}, we find that $\mathrm{Im}\sigma(\omega)=0$ at the zero frequency for values of $\beta$. The real part of the conductivity $\mathrm{Re}\sigma(\omega)$ decreases as $\beta$ increases for all frequency. Particularly, the DC conductivity defined by $\mathrm{Re}\sigma(0)$ approaches zero with the increasing of $\beta$. When $\beta$ is sufficiently large, the DC conductivity vanishes\footnote{The DC conductivity can also be analytically expressed as $\mathrm{Re}\sigma(0)=e^{-\beta \Phi(r_h)}/g_F^2$, as shown in ref.\cite{7}.}, implying that the system behaves effectively like an insulator. Intuitively, we expect that the behavior of the system will be getting closer to an insulator in the increasing process of $\beta$.
\subsection{Conductivity of the superconducting phase}

\begin{figure}
\centering
  \includegraphics[width=5.5cm]{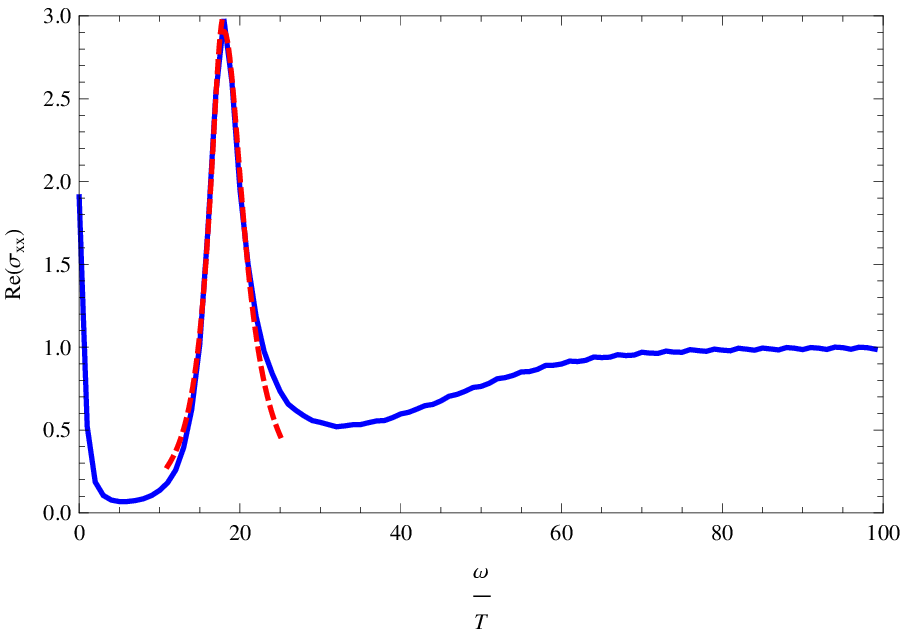}
  \includegraphics[width=5.5cm]{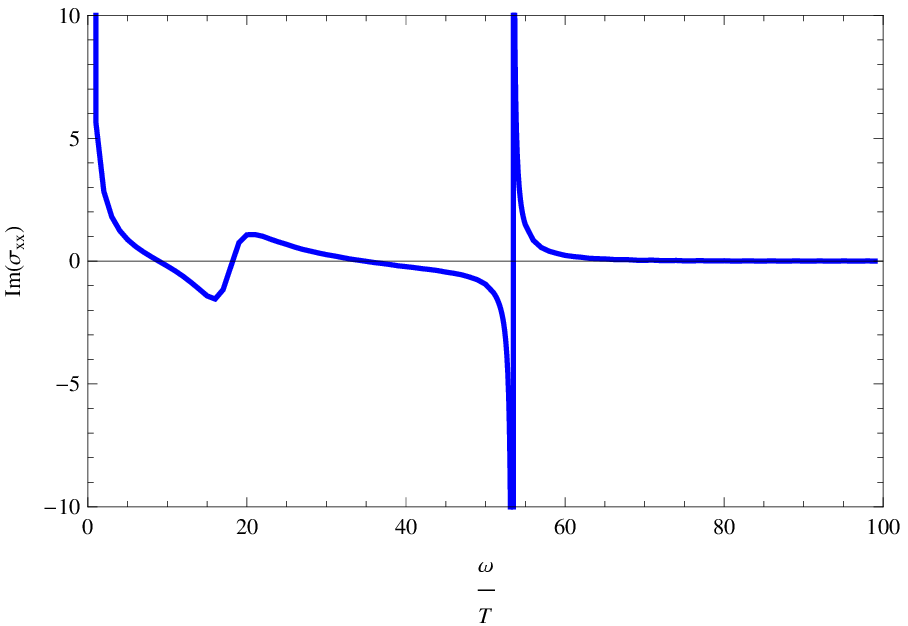}
  \includegraphics[width=5.5cm]{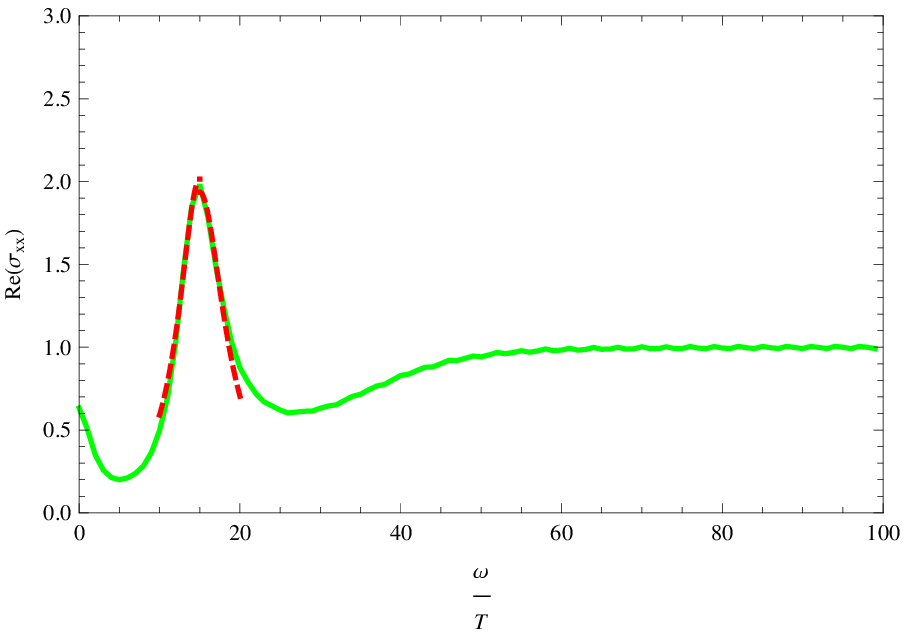}
  \includegraphics[width=5.5cm]{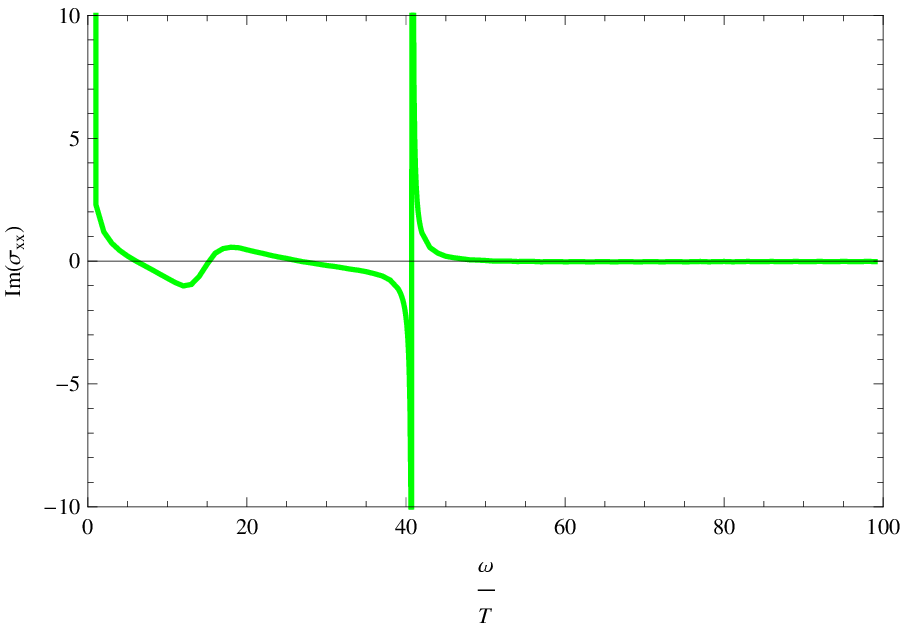}
  \includegraphics[width=5.5cm]{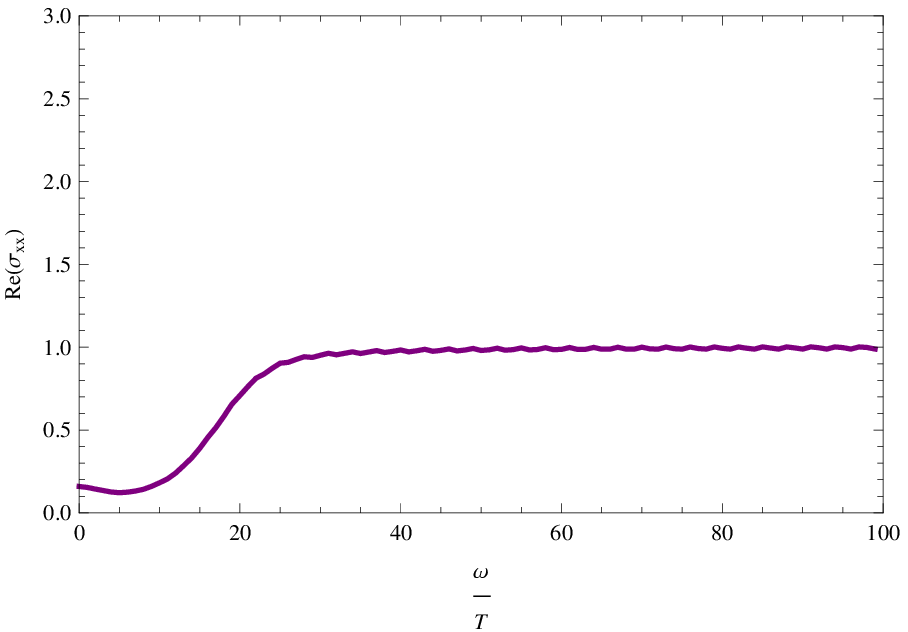}
  \includegraphics[width=5.5cm]{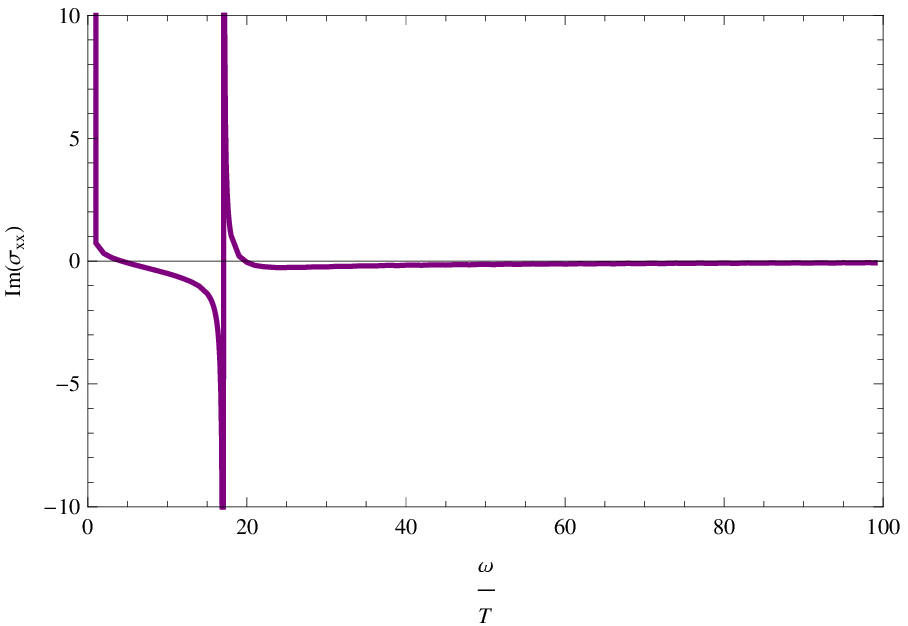}
  \caption{The real (left plots) and imaginary (right plots) part of the conductivity $\sigma_{xx}$ for $\nu=1.4$ and $\mu/\mu_c=1.2$. $\beta=0$ (blue), $\beta=1.5$ (green), $\beta=6$ (purple). The dotted lines are fitted cures.}\label{fig4}
\end{figure}

We are readily to numerically solve the coupled equations of motion eq.(\ref{eom9}-\ref{eom11}) with proper boundary conditions at the horizon. In fig.\ref{fig4}, we present the numerical results of $\sigma_{xx}$ for various value of $\beta$. For small $\beta$, there are some common features in the real and imaginary part of conductivities. For example, when $\beta=0$, there exists a peak at $\omega/T=18$ in the dissipative conductivity which precisely corresponds to a zero point in the imaginary part. This is a characteristic of Drude model. Indeed, we find that the behaviors of the conductivity $\sigma_{xx}$ near the peak can be well approximated by a Drude-like model.
\begin{equation} \sigma_{xx}=\frac{\sigma_0}{1-i(\omega-\omega_0)\tau} \label{drude} \end{equation}
The imaginary part of the conductivity contains two poles: $\omega=0$ and $\omega=\mu$. The pole at $\omega=0$ implies the existence of an infinite DC conductiviy from  Kramers-Kr\"{o}nig relation
\begin{equation} Im[\sigma(\omega)]=-P\  \int_{-\infty}^\infty \frac{\mathrm{d}\omega'}{\pi} \frac{Re[\sigma(\omega)]}{\omega'-\omega} \end{equation}
Near the zero frequency, $Im[\sigma(\omega)]\sim 1/\omega$ which results to $Re[\sigma(\omega)]\sim \pi \delta(\omega)$. The other pole which appears at non-zero frequency is a general consequence of p-wave superconductors (at least in the classical limit) although it is unexpected at the very start\cite{4}.

When we turn on non-zero $\beta$, the maximum height of the peak in the real part of the conductivity begins to be suppressed. The peak even disappears when $\beta$ becomes sufficiently large. From fig.\ref{fig4}, we also find that the non-zero pole moves to a relative small frequency gradually when $\beta$ increases. This is not something surprised since we have fixed $\mu=1.2\mu_c$ while $\mu_c$ decreases for the values of $\beta$ taken in these figures. This is consistent with our observations in fig.\ref{fig2}.

\section{Conclusions}

In this paper, we introduce a dilaton in the bulk to effectively break the scale invariance of the bulk geometry. By coupling the dilaton to the non-Abelian SU(2) gauge field, we study the p-wave superconductors. We find that the conductivity in the normal phase is monotonically decreasing as $\beta$ increases. The DC conductivity approaches zero and vanishes when $\beta$ is sufficiently large. From this perspective, the system can be viewed as an insulator. We expect that it will behave more and more like an insulator in the increasing process of $\beta$. In general, the order parameter condenses when the chemical potential exceeds a critical value, indicating the superconducting transition. The critical potential generally decreases as $\beta$ increases. Furthermore, the Drude peak in the x-direction conductivity $\sigma_{xx}$ is suppressed gradually with the increasing of $\beta$. For $\beta=6$, the peak even gets smoothed out.

\section{Acknowledgments}
I thank Professor Sije. Gao and Dr. Hongbao Zhang for their comments on this manuscript. This work is supported by NSFC Grants NO.  11375026, NO. 11235003 and NCET-12-0054.

\end{document}